\documentclass[prd,amsmath,amssymb,superscriptaddress,showkeys,showpacs,twocolumn,10pt]{revtex4-1}

\pdfoutput=1

\usepackage{graphicx}
\usepackage{dcolumn}
\usepackage{bm}
\usepackage{amssymb}
\usepackage{latexsym}
\usepackage{booktabs}
\usepackage{amsmath}
\usepackage{multirow}
\usepackage{url}
\usepackage{subfigure}

\usepackage{float}
\usepackage[colorlinks=true, linkcolor=red, citecolor=blue]{hyperref}

\begin{document}

\title{Prospects for measuring dark energy with 21 cm intensity mapping experiments: A joint survey strategy}

\author{Peng-Ju Wu}
\affiliation{Key Laboratory of Cosmology and Astrophysics (Liaoning Province) \& Department of Physics, College of Sciences, Northeastern University, Shenyang 110819, China}

\author{Yichao Li}
\affiliation{Key Laboratory of Cosmology and Astrophysics (Liaoning Province) \& Department of Physics, College of Sciences, Northeastern University, Shenyang 110819, China}

\author{Jing-Fei Zhang}
\affiliation{Key Laboratory of Cosmology and Astrophysics (Liaoning Province) \& Department of Physics, College of Sciences, Northeastern University, Shenyang 110819, China}

\author{Xin Zhang}\thanks{Corresponding author.\\zhangxin@mail.neu.edu.cn}
\affiliation{Key Laboratory of Cosmology and Astrophysics (Liaoning Province) \& Department of Physics, College of Sciences, Northeastern University, Shenyang 110819, China}
\affiliation{Key Laboratory of Data Analytics and Optimization for Smart Industry (Ministry of Education), Northeastern University, Shenyang 110819, China}
\affiliation{National Frontiers Science Center for Industrial Intelligence and Systems Optimization, Northeastern University, Shenyang 110819, China}

\begin{abstract}
The 21 cm intensity mapping (IM) technique provides us with an efficient way to observe the cosmic large-scale structure (LSS). From the LSS data, one can use the baryon acoustic oscillation and redshift space distortion to trace the expansion and growth history of the universe, and thus measure the dark energy parameters. In this paper, we make a forecast for cosmological parameter estimation with the synergy of three 21 cm IM experiments. Specifically, we adopt a novel joint survey strategy, FAST\,($0<z<0.35$)\,+\,SKA1-MID\,($0.35<z<0.8$)\,+\,HIRAX\,($0.8<z<2.5$), to measure dark energy. We simulate the 21 cm IM observations under the assumption of excellent foreground removal. We find that the synergy of three experiments could place quite tight constraints on cosmological parameters. For example, it provides $\sigma(\Omega_{\rm m})=0.0039$ and $\sigma(H_0)=0.27\ \rm km\ s^{-1}\ Mpc^{-1}$ in the $\Lambda$CDM model. Notably, the synergy could break the cosmological parameter degeneracies when constraining the dynamical dark energy models. Concretely, the joint observation offers $\sigma(w)=0.019$ in the $w$CDM model, and $\sigma(w_0)=0.085$ and $\sigma(w_a)=0.32$ in the $w_0w_a$CDM model. These results are better than or equal to those given by CMB+BAO+SN. In addition, when the foreground removal efficiency is relatively low, the strategy still performs well. Therefore, the 21 cm IM joint survey strategy is promising and worth pursuing.
\end{abstract}


\maketitle
\section{Introduction}\label{sec1}
The six-parameter $\Lambda$ cold dark matter ($\Lambda$CDM) model is simple and powerful, which fits the cosmic microwave background (CMB) data with stunning precision \citep{Aghanim:2018eyx}. However, the standard model of cosmology has recently shown some cracks. It was found that the CMB results for the $\Lambda$CDM cosmology are in tension with some late-universe observations (see Ref.~\cite{Verde:2019ivm} for a review). In addition, $\Lambda$CDM has some theoretical problems, such as the ``fine-tuning'' and ``cosmic coincidence'' problems \citep{Weinberg:1988cp,Sahni:1999gb}. All these imply that the $\Lambda$CDM model needs to be further extended. Cosmologists have conceived a variety of theories beyond the $\Lambda$CDM model to reconcile the tensions and solve the problems \cite{Guo:2018ans}. However, as an early-universe probe, CMB cannot tightly constrain the newly introduced parameters related to the late-time physics, such as the equation-of-state (EoS) parameters of dark energy \citep{Aghanim:2018eyx}. The large constraint errors leave the possibility for various dynamical dark energy models.

In order to tightly constrain these models or exclude them with high confidence, one should use late-universe probes to precisely measure the evolution of the universe. Baryon acoustic oscillations (BAOs) are frozen sound waves originating from the photon-baryon plasma prior to recombination, which leave an imprint on the large-scale structure (LSS) in the universe at a characteristic scale of $r_{\rm d} \simeq 147\ \rm Mpc$ \citep{Aghanim:2018eyx}. The BAO scale can be used as a standard ruler to measure the angular diameter distance $D_{\rm A}(z)$ and Hubble parameter $H(z)$, and hence serves as a useful tool to constrain the cosmological parameters. By using galaxy redshift surveys to map the distribution of matter, one can extract the BAO features in the power spectrum \citep{Seo:2003pu,Blake:2003rh,Weinberg:2013agg}. In addition, one can derive the structure growth rate $f(z)$ from the redshift space distortions (RSDs), which is also useful for cosmological parameter estimation. Currently, the BAO observations cannot tightly constrain the cosmological parameters. It should be pointed out that the galaxy redshift survey is a time-consuming process that requires resolving individual galaxies. In the next decades, a promising way to measure the BAO and RSD signals is the 21 cm intensity mapping (IM) method.

After the epoch of reionization (EoR), most of the neutral hydrogen (\textsc{H\,i}) is believed to exist in dense gas clouds within galaxies \citep{Barkana:2006ep}. \textsc{H\,i} is therefore a tracer of the galaxy distribution, and thus the overall matter distribution. By measuring the total \textsc{H\,i} 21 cm emission of many galaxies within large voxels, one can also obtain the LSS map of the universe. The technique allows us to measure the LSS more efficiently. The detection of 21 cm signal in the IM regime has been tested with existing telescopes, and several cross-correlation power spectra between 21 cm IM maps and galaxy maps have been detected~\citep{Chang:2010jp,Masui:2012zc, Anderson:2017ert, CHIME:2022kvg,Cunnington:2022uzo}. But until now, no experiment has detected the 21 cm IM power spectrum in auto-correlation. However, the situation will be changed with the advent of some instruments, such as BINGO~\cite{Abdalla:2021nyj}, FAST~\cite{Nan:2011um,Jing2021}, MeerKAT \citep{MeerKLASS:2017vgf,Wang:2020lkn}, SKA1-MID~\cite{Santos:2015gra,Xu:2020uws,An2022}, HIRAX~\cite{Crichton:2021hlc}, CHIME~\cite{CHIME:2022dwe}, and Tianlai \citep{Chen:2012xu, Li:2020ast}. Among them, FAST, MeerKAT, CHIME, and Tianlai (pathfinder array) are already in operation. More recently, CHIME detected the \textsc{H\,i} cross-correlation with the luminous red galaxies at $z=0.84$, emission line galaxies at $z=0.96$, and quasars at $z=1.20$ \citep{CHIME:2022kvg}. In addition, the MeerKAT team presented a $7.7\sigma$ detection of the cross-correlation with WiggleZ galaxies \citep{Cunnington:2022uzo}.

The performance of 21 cm IM experiments in cosmological parameter constraints has been widely discussed. For example, the 21 cm observations can be used to measure the nature of dark energy \citep{Bull:2014rha, Xu:2014bya, Yahya:2014yva, Villaescusa-Navarro:2015cca, Pourtsidou:2016dzn, Witzemann:2017lhi, Xu:2017rfo, Olivari:2017bfv, Santos:2017bqm, Yohana:2019ahg, Zhang:2019ipd, Xu:2020uws, Zhang:2021yof, Wu:2021vfz, Berti:2021ccw, Jin:2021pcv, Wu:2022dgy, Berti:2022ilk, Karagiannis:2022ylq,Zhang:2023gaz}, the inflationary features in the primordial power spectrum \citep{Xu:2016kwz,CosmicVisions21cm:2018rfq}, and the primordial non-Gaussianity \citep{Tashiro:2012wr,DAloisio:2013mgn,Camera:2014bwa,Xu:2014bya,Li:2017jnt,Karagiannis:2019jjx}, as well as to test the hemispherical power asymmetry \citep{Shiraishi:2016omb,Li:2019bsg}. In recent years, combining the 21 cm experiments with other observations to improve the cosmological parameter estimation has been studied \citep{Jin:2020hmc,Jin:2021pcv,Zhang:2021yof,Wu:2022dgy}, but the synergy of multiple 21 cm experiments was rarely mentioned. In this paper, we explore the prospects of using a novel 21 cm IM joint survey strategy, FAST\,($0<z<0.35$)\,+\,SKA1-MID\,($0.35<z<0.8$)\,+\,HIRAX\,($0.8<z<2.5$), to measure dark energy. We shall simulate the 21 cm IM observations under the assumption of excellent foreground subtraction, and use the mock data to constrain three typical dark energy models, namely, the $\Lambda$CDM, $w$CDM, and $w_0w_a$CDM models. We note that SKA1-MID could perform the 21 cm IM observations in the redshift interval $0.35<z<3.05$, but we only consider its survey at $0.35<z<0.8$ in the joint observation. To illustrate the potential of our survey strategy, we shall compare the constraint results with those given by HIRAX alone (which plays a dominant role in the joint observation) and also with those given by the combination of three mainstream observations, CMB+BAO+SN (here, BAO refers to those measured form galaxy redshift surveys, and SN refers to the observations of type Ia supernovae).

The remainder of this paper is organized as follows. We briefly describe the methodology in Sec.~\ref{sec2}. In Sec.~\ref{sec3}, we present the results and make some discussions. Finally, we give our conclusions in Sec.~\ref{sec4}.

\section{methodology}\label{sec2}
In this work, we employ the flat $\Lambda$CDM model with $H_0=67.3\ \rm km\ s^{-1}\ Mpc^{-1}$, $\Omega_{\rm m}=0.317$, and $\sigma_8=0.812$ as a fiducial cosmology to generate the mock data.

In the 21 cm IM regime, the location of an observed pixel is given by \cite{Bull:2014rha},
\begin{align}
{\boldsymbol{r}}_{\perp} &= D_{\rm C}(z_i)\big({\boldsymbol{\theta}_{p}}-{\boldsymbol{\theta}}_{i}\big),\ r_{\parallel} = D_{\nu}(z_{i})\big(\tilde{\nu}_{p}-\tilde{\nu}_{i}\big),
\end{align}
where ${\boldsymbol{\theta}_{p}}$ is the angular direction, ${\nu}_{p}$ is the frequency, and we have centered the survey on $\big({\boldsymbol{\theta}}_{i},{\nu}_{i}\big)$, corresponding to a redshift bin centered at $z_{i}$. $D_{\rm C}(z)$ is the comoving distance and $D_{\nu}(z)$ is defined as $c(1+z)^2/H(z)$, with $c$ the speed of light. $\tilde{\nu} \equiv \nu/\nu_{21}$, with $\nu_{21}=1420.4\ \rm MHz$ the frequency of 21-cm line. We adopt the observational coordinates $(\boldsymbol{q}={\boldsymbol{k}}_{\perp}D_{\rm C}, y=k_{\parallel}D_{\nu})$, where ${\boldsymbol{k}}_{\perp}$ and $k_{\parallel}$ are the perpendicular and parallel components of the wave vector $\boldsymbol{k}$, respectively.

For a clump of \textsc{H\,i}, the mean brightness temperature can be written as \citep{Hall:2012wd}
\begin{align}
\overline{T}_{\rm b}(z)=188h\Omega_{\textsc{H\,i}}(z)\displaystyle{\frac{(1+z)^2}{E(z)}\,\rm mK},
\end{align}
where $\Omega_{\textsc{H\,i}}(z)$ is the \textsc{H\,i} fractional density for which we use the form shown in Ref.~\citep{Bull:2014rha}, $h$ is the dimensionless Hubble constant, and $E(z)\equiv H(z)/H_0$ is the reduced Hubble parameter. Considering the RSDs, the 21\,cm signal covariance is given by \citep{Bull:2014rha}
\begin{align}
C^{\rm S}(\boldsymbol{q},y)=&\displaystyle{\frac{\overline{T}_{\rm b}^{2}(z_i)\alpha_{\perp}^2\alpha_{\parallel}}{D_{\rm C}^2D_{\nu}}}\left(b_{\textsc{H\,i}}+f\mu^2\right)^2\exp{\left(-k^2 \mu^2 \sigma_{\rm NL}^{2}\right)} \nonumber \\
&\times D^2(z_i)P(k,z=0),
\end{align}
where $\alpha_{\perp}\equiv D_{\rm A}^{\rm fid}(z)/D_{\rm A}(z)$ and $\alpha_{\parallel}\equiv H(z)/H^{\rm fid}(z)$, and the superscript ``fid'' stands for the fiducial cosmology. $b_{\textsc{H\,i}}$ refers to the \textsc{H\,i} bias, $f(z)\simeq\Omega_{\rm m}^{\gamma}(z)$ is the linear growth rate with $\gamma=0.545$ for $\Lambda$CDM, $\mu\equiv k_{\parallel}/k$, $\sigma_{\rm NL}$ is the non-linear dispersion scale, and $P(k,z=0)$ is the the present matter power spectrum generated by {\tt CAMB} \citep{Lewis:1999bs}. In this paper, we adopt the Planck best-fit primordial power spectrum to determine the inflationary parameters in $P(k)$. $D(z)$ is the linear growth factor which is related to $f(z)$ by
\begin{align}
f(z)=-\displaystyle{\frac{1+z}{D(z)}}\displaystyle{\frac{{\rm d} D(z)}{{\rm d} z}}.
\end{align}

We now turn to noises and effective beams. The noise covariance is given by \citep{Bull:2014rha}
\begin{align}
C^{\rm N}(\boldsymbol{q},y)=\displaystyle{\frac{\sigma_{\rm pix}^2V_{\rm pix}}{D_{\rm C}^2D_{\nu}}} B_{\parallel}^{-1} B_{\perp}^{-2},
\end{align}
where $\sigma_{\rm pix}$ is the pixel noise, $V_{\rm pix}=D_{\rm C}^2{\rm FoV}\times D_{\nu}\delta\nu/\nu_{21}$ is the pixel volume, with $\rm FoV$ the field of view of receiver and $\delta\nu$ the channel bandwidth. The factors $B_{\parallel}$ and $B_{\perp}$ describe the frequency and angular responses of the instrument, respectively. In this work, we simulate the 21 cm IM data based on the hypothetical observations of FAST, SKA1-MID, and HIRAX, respectively. The experimental configurations are shown in Table~\ref{Telescope}.
\begin{table}
\renewcommand\arraystretch{1.2}
\caption{Experimental configurations for FAST, SKA1-MID and HIRAX.}
\label{Telescope}
\centering
\begin{tabular}{p{1.7cm}|p{1.8cm}<{\centering} p{1.8cm}<{\centering} p{1.8cm}<{\centering}}
\bottomrule[1pt]
                                    & FAST     & SKA1-MID                   & HIRAX        \\
\hline
$z_{\rm min}$                       & 0        & 0.35                       & 0.8          \\
$z_{\rm max}$                       & 0.35     & 3.05                       & 2.5          \\
$N_{\rm d}$                         & 1        & 197                        & 1024         \\
$N_{\rm b}$                         & 19       & 1                          & 1            \\
$D_{\rm d}~[\rm m]$                 & 300      & 15                         & 6            \\
$S_{\rm area}~[\rm{deg^2}]$         & 20,000   & 20,000                    & 15,000        \\
$t_{\rm tot}~[\rm h]$               & 10,000   & 10,000                    & 10,000        \\
$T_{\rm rec}~[\rm K]$               & 20       & Eq.~(\ref{SKA1-MID})       & 50           \\
\bottomrule[1pt]
\end{tabular}
\end{table}
Note that FAST and SKA1-MID operate in the single-dish mode, while HIRAX operates in the interferometric mode. For an experiment using the single-dish mode,
\begin{align}
\sigma_{\rm pix}=\displaystyle{\frac{T_{\rm sys}}{\sqrt{n_{\rm pol}t_{\rm tot}\delta\nu\left(\rm{FoV}/S_{\rm area}\right)}}}\displaystyle{\frac{\lambda^2}{A_{\rm e}\rm{FoV}}}\displaystyle{\frac{1}{\sqrt{N_{\rm d}N_{\rm b}}}},
\end{align}
and for an interferometer,
\begin{align}
\sigma_{\rm pix}=\displaystyle{\frac{T_{\rm sys}}{\sqrt{n_{\rm pol}t_{\rm tot}\delta\nu\left(\rm{FoV}/S_{\rm area}\right)}}}
\displaystyle{\frac{\lambda^2}{A_{\rm e}\sqrt{\rm FoV}}}
\displaystyle{\frac{1}{\sqrt{n(\boldsymbol{u})N_{\rm b}}}},
\end{align}
where $T_{\rm sys}$ is the system temperature, $t_{\rm tot}$ is the total observing time, $S_{\rm area}$ is the survey area, $n_{\rm pol}=2$ is the number of polarization channels, $A_{\rm e}$ is the effective collecting area of each receiver, $N_{\rm d}$ is the number of dishes and $N_{\rm b}$ is the number of beams. For the dish reflector, $A_{\rm e}=\eta\pi(D_{\rm d}/2)^2$ and $\rm{FoV}\approx\theta_{\rm B}^2$, where $D_{\rm d}$ is the diameter of the dish, $\eta$ is the efficiency factor for which we adopt 0.7, $\theta_{\rm B}\approx\lambda/D_{\rm d}$ is the full width at half-maximum of the beam, and $n(\boldsymbol{u})$ is the baseline density for the interferometer. The system temperature can be refined into three parts,
\begin{align}
T_{\rm sys}=T_{\rm rec} + T_{\rm gal} + T_{\rm CMB},
\end{align}
where $T_{\rm rec}$ is the receiver temperature, $T_{\rm gal}\simeq25\ {\rm K}\times(408\ \rm{MHz}/\nu)^{2.75}$ is the contribution from the Milky Way, and $T_{\rm CMB}\simeq2.725\ {\rm K}$ is the CMB temperature. For SKA1-MID, $T_{\rm rec}$ is assumed to be \citep{SKA:2018ckk}
\begin{align}
\label{SKA1-MID}
T_{\rm rec}=15\ {\rm K}+ 30\ {\rm K}\left(\nu/{\rm GHz}-0.75\right)^2.
\end{align}

The 21 cm signal is also contaminated by astrophysical foregrounds which are much brighter than the signal. However, the foregrounds have a smooth spectral shape, so in principle one can use some algorithms to remove them \cite{deOliveira-Costa:2008cxd, Bonaldi:2014zma, Mertens:2017gxw,Hothi:2020dgq,Soares:2021ohm,Ni:2022kxn,Gao:2022xdb}. In this paper, we assume that a removal algorithm has been applied and only consider the effect of residual foreground. The covariance of residual foreground is calculated by \citep{Bull:2014rha}
\begin{align}
C^{\rm F}(\boldsymbol{q},y)=\varepsilon_{\rm FG}^2\sum_{X}A_{X}\left(\displaystyle{\frac{l_p}{2\pi q}}\right)^{n_{X}}\left(\displaystyle{\frac{\nu_p}{\nu_i}}\right)^{m_{X}},
\end{align}
where $\varepsilon_{\rm FG}$ characterizes the foreground removal efficiency: $\varepsilon_{\rm FG}=1$ corresponds to no removal and $\varepsilon_{\rm FG}=0$ corresponds to perfect removal. Unless otherwise specified, we consider an optimistic scenario of $\varepsilon_{\rm FG}=10^{-6}$. For a foreground $X$, the amplitude ($A_X$) and index ($n_X$ and $m_X$) parameters at $l_p=1000$ and $\nu_p=130\ \rm MHz$ can be found in Ref.~\cite{Santos:2004ju}.

The total covariance is $C^{\rm T}=C^{\rm S}+C^{\rm N}+C^{\rm F}$, and the Fisher matrix for a set of parameters $\{p\}$ in a redshift bin is given by \citep{Bull:2014rha}
\begin{align}
{F}_{ij}=\displaystyle{\frac{1}{8\pi^2}}V_{\rm bin}\int_{-1}^{1}{\rm d}\mu\int_{k_{\rm min}}^{k_{\rm max}}k^2{\rm d}k \displaystyle{\frac{\partial\ln{C^{\rm T}}}{\partial p_{i}}} \displaystyle{\frac{\partial\ln{C^{\rm T}}}{\partial p_{j}}},
\end{align}
where $V_{\rm bin}$ is the survey volume. We assume that $b_{\textsc{H\,i}}$ is only redshift dependent (appropriate for large scales), so a non-linear cut-off at $k_{\rm max}=0.14(1+z)^{2/3}$ is imposed \citep{Smith:2002dz}. In addition, the largest scale that the survey can probe corresponds to a wave vector $k_{\rm min}=2\pi V_{\rm bin}^{-1/3}$. We simulate the measurement of 21 cm IM power spectrum and calculate the Fisher matrix for $\{p\}$ in each redshift bin ($\Delta z=0.1$). The parameter set \{p\} is selected as $\{D_{\rm A}(z), H(z), [f\sigma_8](z), [b_{\textsc{H\,i}}\sigma_8](z), \sigma_{\rm NL}\}$ in this work. Note that we marginalize $[b_{\textsc{H\,i}}\sigma_8](z)$ and $\sigma_{\rm NL}$, and only use the Fisher matrix of $\{D_{\rm A}(z), H(z), [f\sigma_8](z)\}$ to constrain cosmological parameters. The inverse of Fisher matrix can provide measurement errors on $D_{\rm A}(z)$, $H(z)$, and $[f\sigma_8](z)$.

\begin{figure}
\includegraphics[scale=0.5]{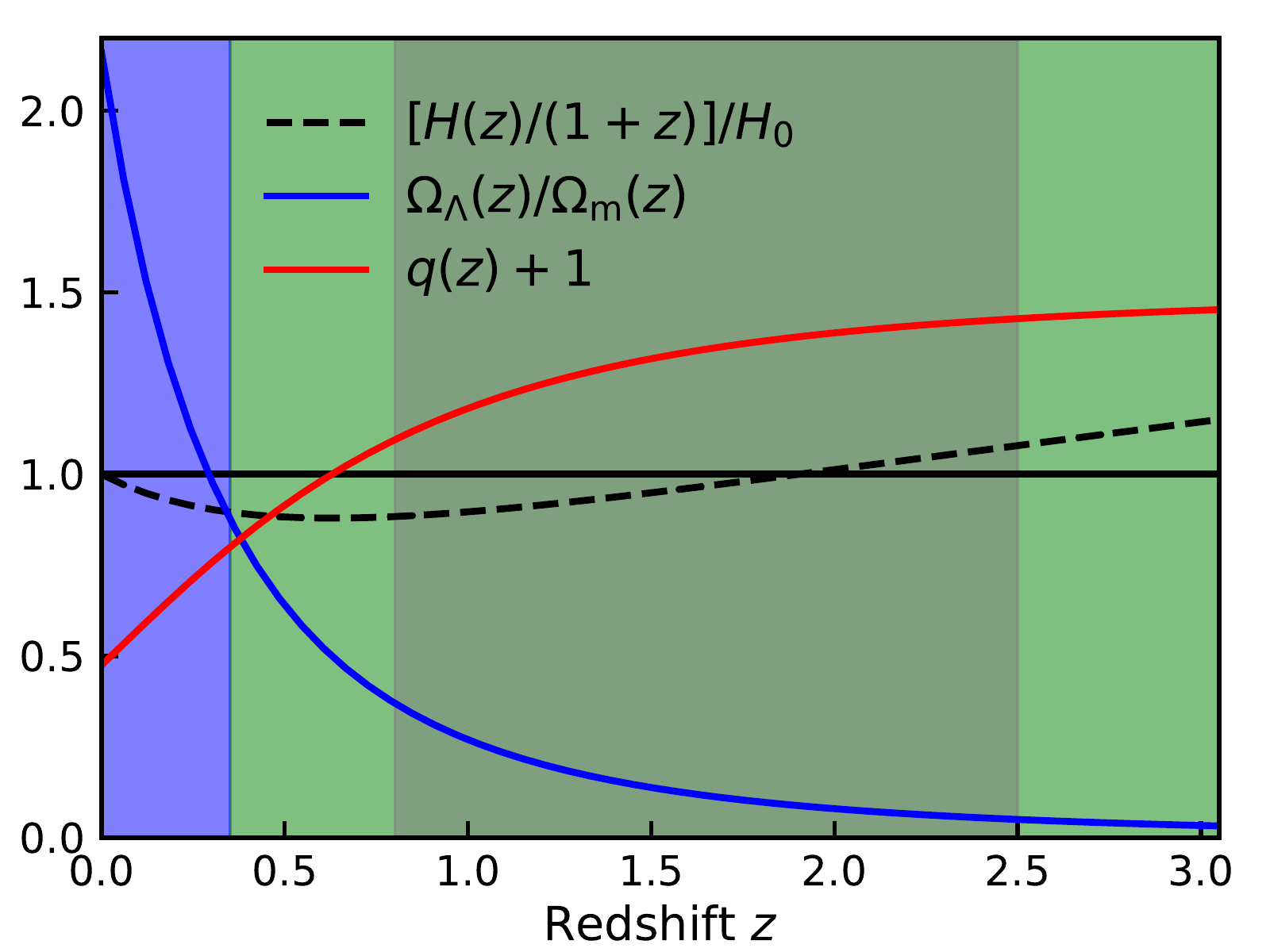}
\centering
\caption{Schematic view of the redshift coverage for FAST (blue shaded region), SKA1-MID (green shaded region), and HIRAX (grey shaded region). Also shown are the redshift evolutions of the functions $\Omega_{\Lambda}(z) / \Omega_{\rm m}(z)$, $[H(z)/(1+z)]/H_0$, and $q(z)+1$. Here $q(z)$ is the deceleration parameter.}
\label{fig:view}
\end{figure}

\begin{figure*}[!htbp]
\includegraphics[scale=0.37]{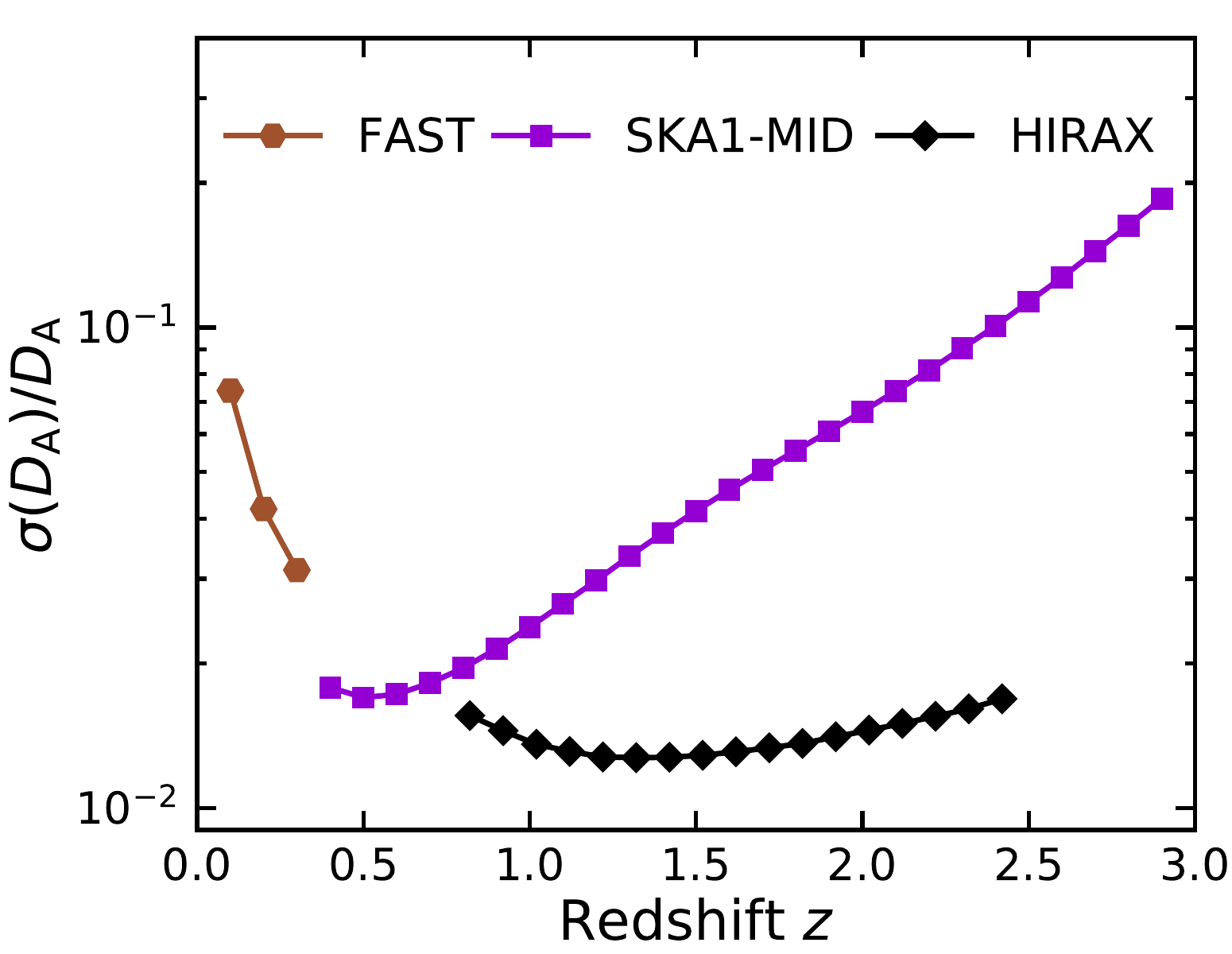}
\includegraphics[scale=0.37]{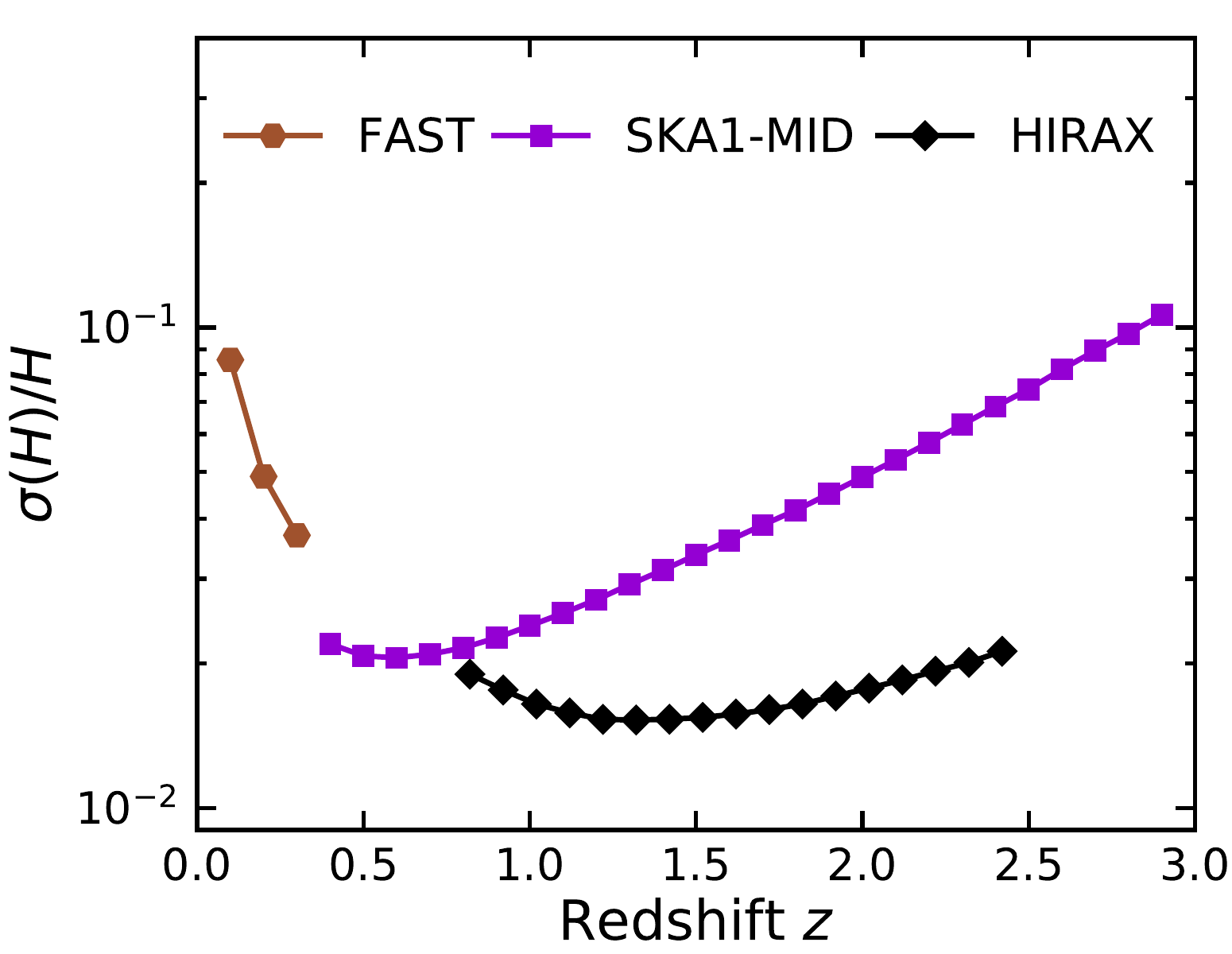}
\includegraphics[scale=0.37]{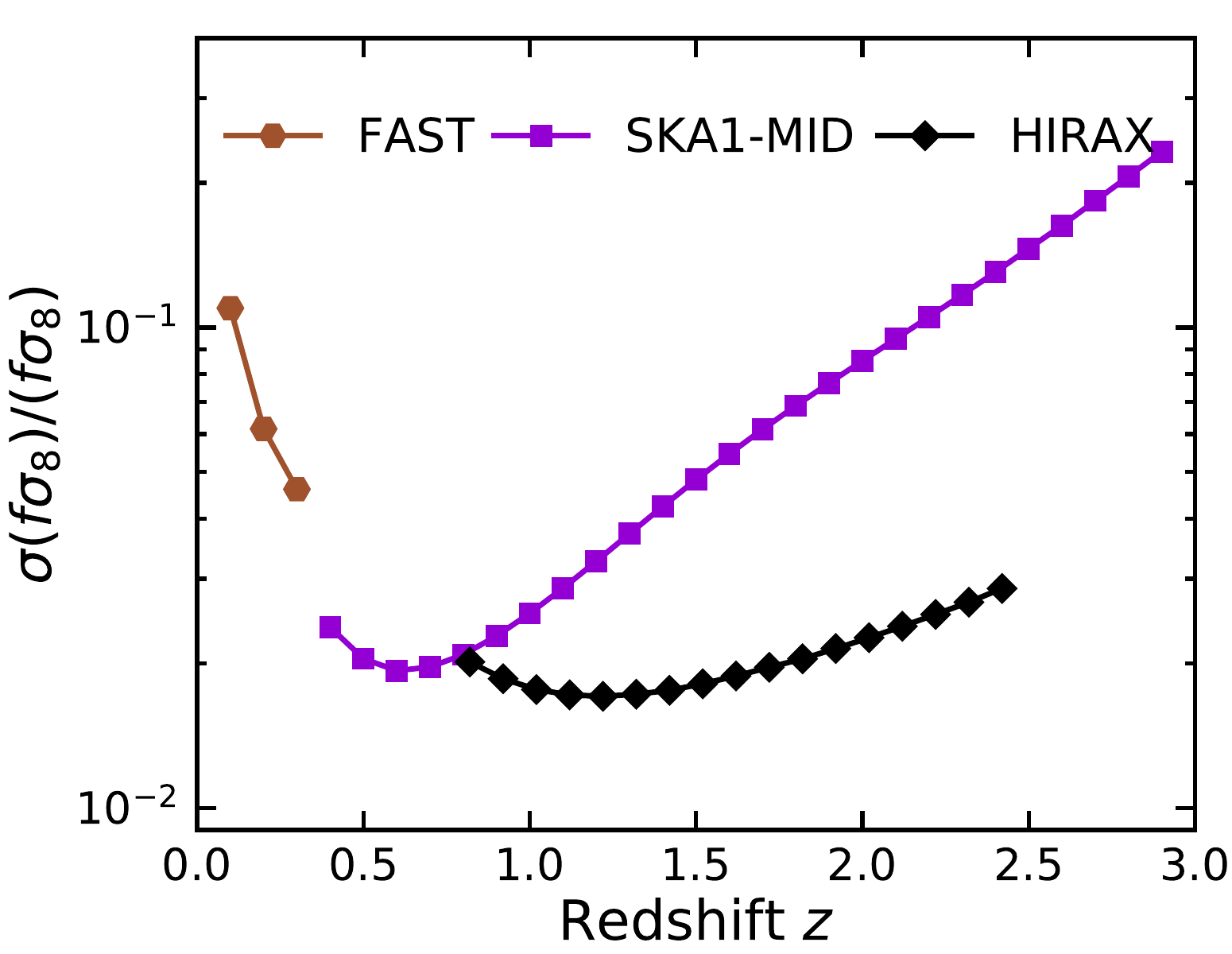}
\caption{Measurement errors on $D_{\rm A}(z)$ (left panel), $H(z)$ (central panel), and $[f\sigma_8](z)$ (right panel) of FAST, SKA1-MID, and HIRAX, in the case of $\varepsilon_{\rm FG}=10^{-6}$. Note that for SKA1-MID we only consider the survey at $0.35<z<0.8$ in the joint observation.}
\centering
\label{DAHzfs8}
\end{figure*}

Figure~\ref{fig:view} shows the redshift coverage for FAST, SKA1-MID, and HIRAX. Also shown are the redshift evolutions of three cosmological functions in the Planck best-fit $\Lambda$CDM model, including $\Omega_{\Lambda}(z) / \Omega_{\rm m}(z)$ (the ratio of dark energy density parameter to matter density parameter), $[H(z)/(1+z)]/H_0$, and $q(z)+1$ where $q(z)=-1+(1+z) \frac{d \ln \left[H(z) / H_{0}\right]}{d z}$ is the deceleration parameter. We can see that FAST mainly covers the dark energy-dominated era of the universe, while HIRAX is dedicated to observing the matter-dominated epoch of the universe. SKA1-MID has the largest redshift coverage, and importantly, its survey at $0.35<z<0.8$ contains the transition redshift $z_t$ (which determines the onset of cosmic acceleration).
The measurement errors on $D_{\rm A}(z)$, $H(z)$, and $[f\sigma_8](z)$ for the three experiments are shown in Fig.~\ref{DAHzfs8}. As can be seen, the errors of SKA1-MID grow rapidly at high redshifts, even exceeding 10\% at $z=3$. In contrast, the errors of HIRAX are significantly smaller in the redshift interval $0.8<z<2.5$. In order to give full play to the advantages of three experiments, we adopt a novel joint survey strategy, i.e., FAST\,($0<z<0.35$)\,+\,SKA1-MID\,($0.35<z<0.8$)\,+\,HIRAX\,($0.8<z<2.5$), to measure dark energy.

Having obtained the Fisher matrices for $\{D_{\rm A}(z_n)$, $H(z_n)$, $[f\sigma_8](z_n); n=1,...,N\}$ in $N$ redshift bins, we adopt the Markov Chain Monte Carlo (MCMC) method to maximize the likelihood $\mathcal{L}\propto \exp{(-\chi^2/2)}$ to infer the posterior probability distributions of cosmological parameters. For a redshift bin, the $\chi^2$ function can be written as
\begin{align}
\chi^2 = \sum_{ij}{x}_i{F}_{ij}{x}_j,
\end{align}
where $\boldsymbol{x}=(H^{\rm th}-H^{\rm obs},\, D_{{\rm A}}^{\rm th}-D_{{\rm A}}^{\rm obs},\, {[f\sigma_8]}^{\rm th}-{[f\sigma_8]}^{\rm obs})$. Note that the total $\chi^2$ function is the sum of the $\chi^2$ functions for each redshift bin. To illustrate the potential of our survey strategy, we shall compare the constraint results with those given by HIRAX alone and those achieved by CMB+BAO+SN. For the CMB data, we adopt the Planck 2018 TT,TE,EE+lowE \citep{Aghanim:2018eyx}. For the BAO data, we consider the measurements from SDSS-MGS \citep{Ross:2014qpa}, 6dFGS \citep{Beutler:2011hx}, and BOSS DR12 \citep{BOSS:2016wmc}. For the SN data, we adopt the latest Pantheon sample \citep{Pan-STARRS1:2017jku}.

\section{Results and discussions}\label{sec3}
\begin{figure*}
\includegraphics[scale=0.37]{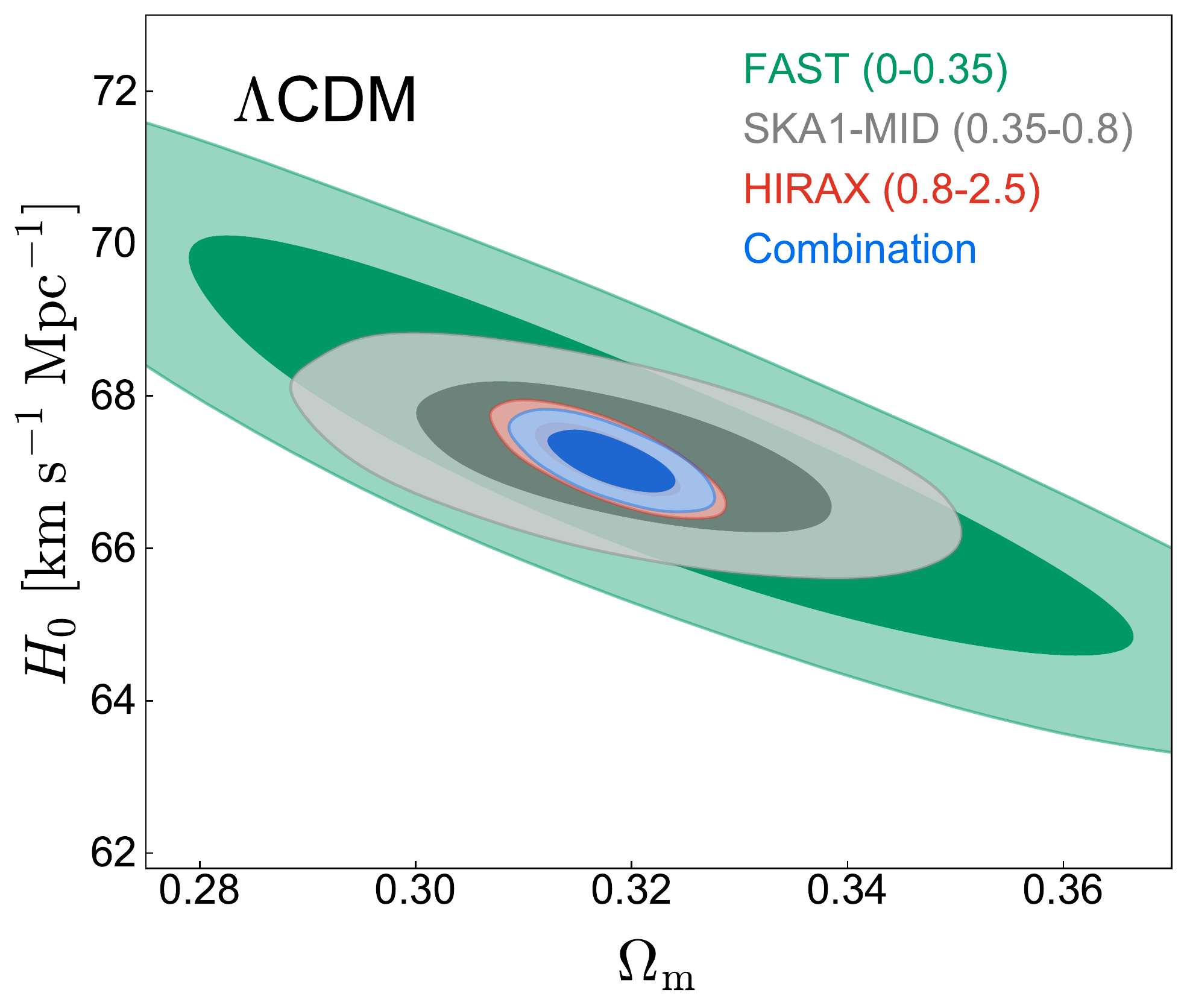}
\includegraphics[scale=0.37]{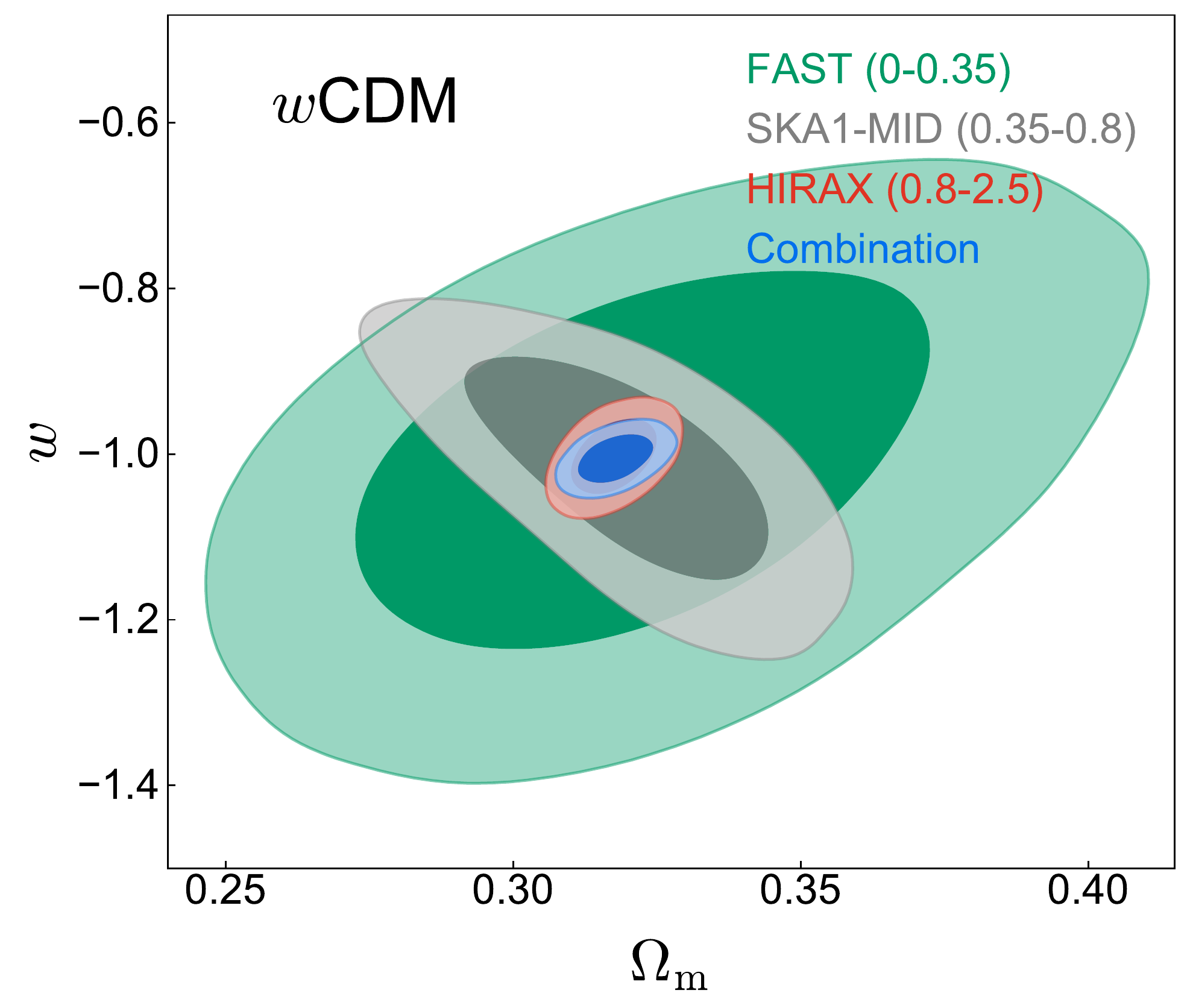}
\includegraphics[scale=0.37]{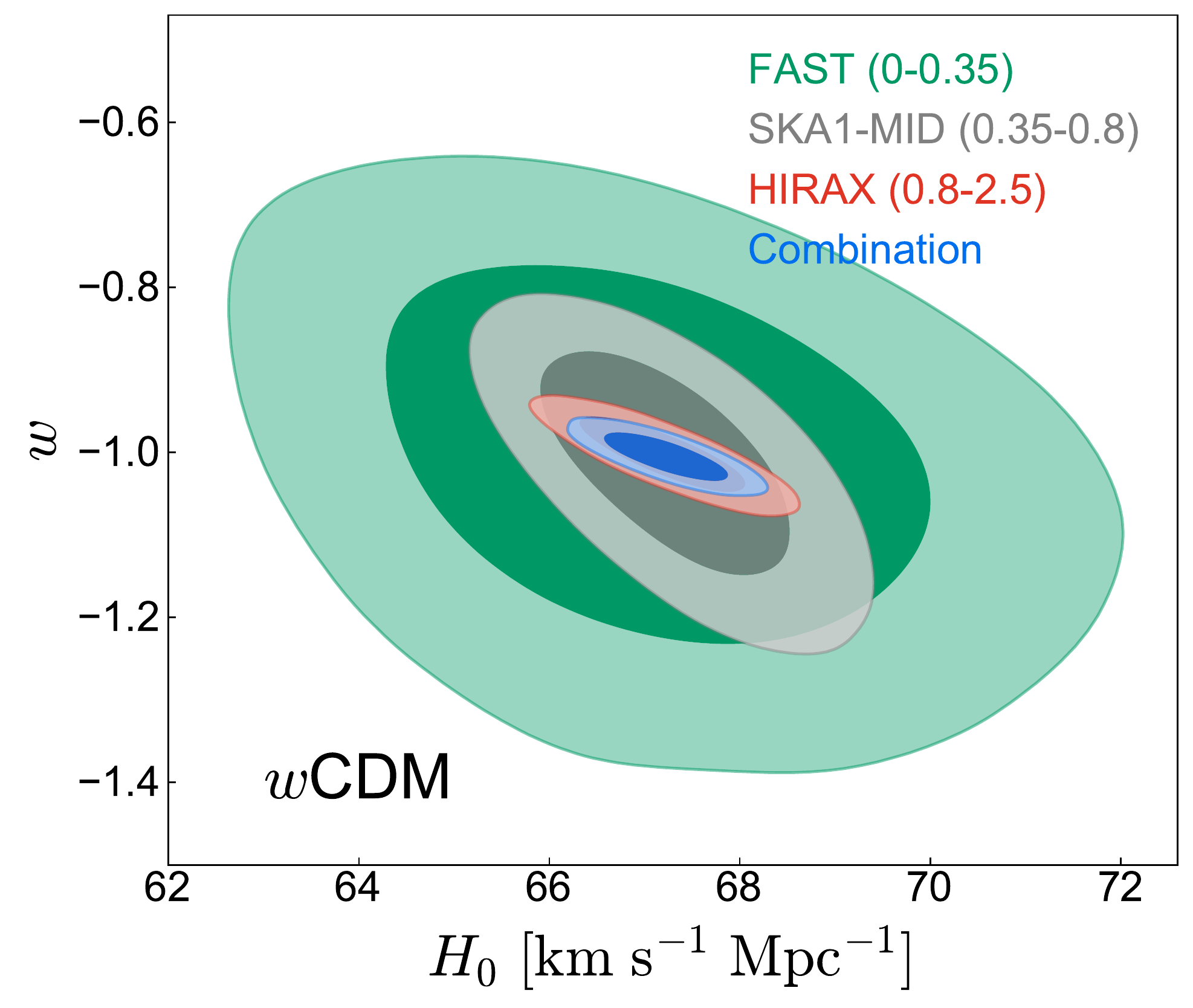}
\includegraphics[scale=0.37]{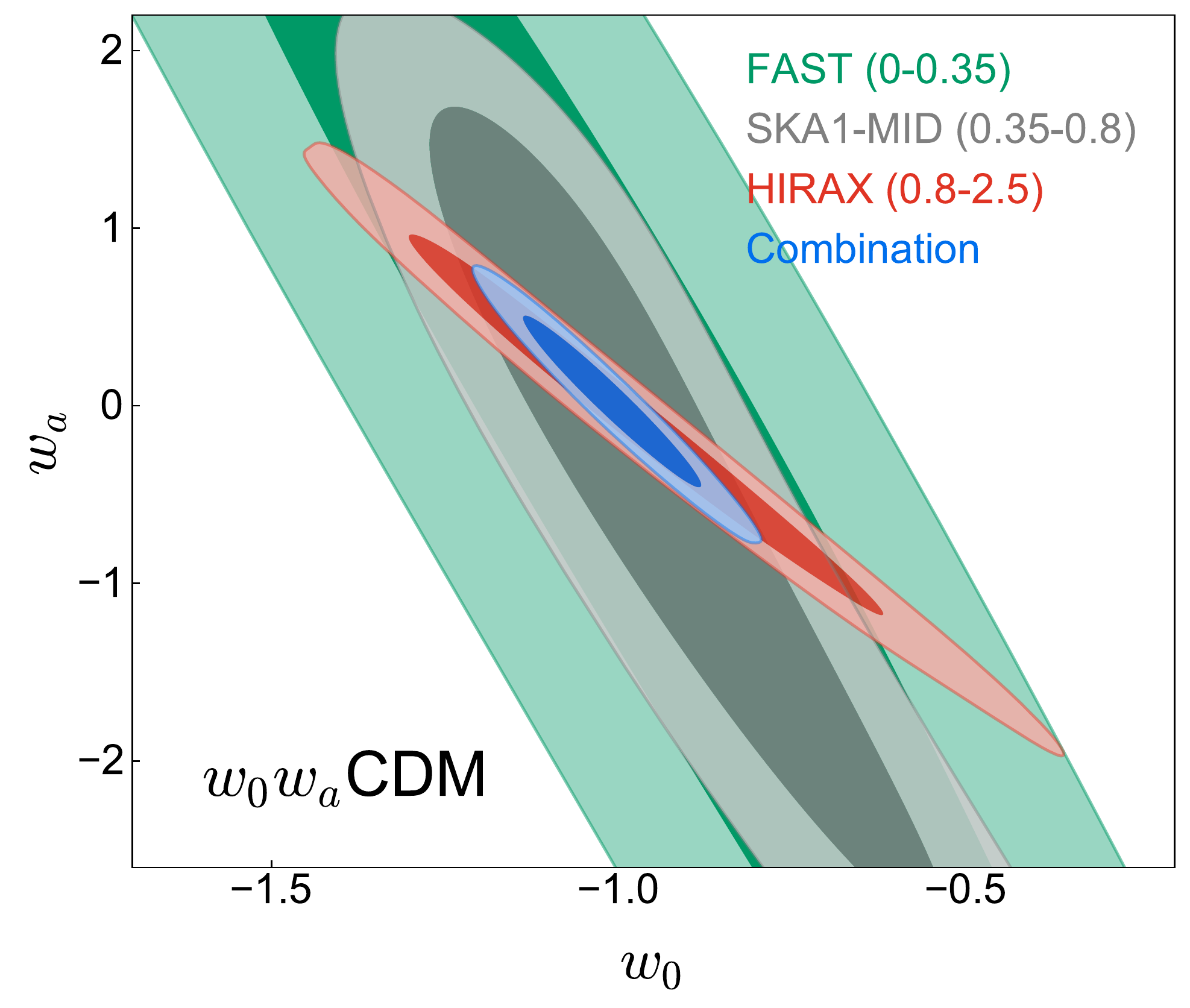}
\centering
\caption{Constraints (68.3\% and 95.4\% confidence level) on the $\Lambda$CDM, $w$CDM, and $w_0w_a$CDM models, by using the FAST, SKA1-MID\,(0.35\,--\,0.8), HIRAX, and FAST+SKA1-MID\,(0.35\,--\,0.8)+HIRAX data, in the case of $\varepsilon_{\rm FG}=10^{-6}$.}
\label{fig:figure}
\end{figure*}

In this section, we report the constraint results. Here we consider three typical cosmological models: (\romannumeral1) $\Lambda$CDM model---the standard cosmological model with $w(z)=-1$; (\romannumeral2) $w$CDM model---the dynamical dark energy model with a constant EoS $w(z)=w$; (\romannumeral3) $w_0w_a$CDM model---the dynamical dark energy model with an evolving EoS $w(z)=w_0+w_a z/(1+z)$ \citep{Chevallier:2000qy,Linder:2002et}. The cosmological parameters we sample include $H_0$, $\Omega_{\rm m}$, $\sigma_8$, $w$, $w_0$, and $w_a$, and we take flat priors for them. The $1\sigma$ and $2\sigma$ posterior distribution contours for various cosmological parameters of interest are shown in Fig.~\ref{fig:figure}, and the $1\sigma$ errors for the marginalized parameter constraints are summarized in Table~\ref{tab:result}. For convenience, we use SKA1-MID\,(0.35\,--\,0.8) to represent the survey of SKA1-MID at $0.35<z<0.8$, and FSH and CBS to represent FAST+SKA1-MID\,(0.35\,--\,0.8)+HIRAX and CMB+BAO+SN respectively. Unless otherwise specified, SKA1-MID refers to the survey of SKA1-MID at $0.35<z<3.05$. In the following discussions, we shall use $\sigma(\xi)$ to represent the $1\sigma$ absolute error of the parameter $\xi$.

\begin{table*}[!htb]
\caption{The 1$\sigma$ errors on the cosmological parameters in the $\Lambda$CDM, $w$CDM, and $w_0w_a$CDM models, by using the FAST, SKA1-MID\,(0.35\,--\,0.8), SKA1-MID, HIRAX, FAST+SKA1-MID\,(0.35\,--\,0.8)+HIRAX (FSH), and CMB+BAO+SN (CBS) data. Here $H_0$ is in units of $\rm km\ s^{-1}\ Mpc^{-1}$.}
\label{tab:result}
\setlength{\tabcolsep}{0.2mm}
\renewcommand{\arraystretch}{1.2}
\begin{center}{\centerline{
\begin{tabular}{ccm{1.5cm}<{\centering}m{1.7cm}<{\centering}m{1.7cm}<{\centering}m{1.5cm}<{\centering}m{1.3cm}<{\centering}m{1.3cm}<{\centering}}
\hline
Model       & Error                         &FAST           &SKA1-MID (0.35--0.8) &SKA1-MID         &HIRAX          &FSH              &CBS            \\ \hline
 \multirow{2}{*}{$\Lambda$CDM}
            &$\sigma(\Omega_{\rm m})$       &$0.029$        &$0.013$             &$0.0066$         &$0.0044$       &$0.0039$         &$0.006$        \\
            &$\sigma(H_0)$                  &$1.8$          &$0.65$             &$0.52$           &$0.32$         &$0.27$           &$0.44$         \\ \hline
 \multirow{3}{*}{$w$CDM}
            &$\sigma(\Omega_{\rm m})$       &$0.032$        &$0.017$             &$0.0069$          &$0.0049$       &$0.0043$         &$0.0076$       \\
            &$\sigma(H_0)$                  &$1.9$          &$0.87$             &$0.67$           &$0.58$         &$0.43$           &$0.82$         \\
            &$\sigma(w)$                    &$0.15$         &$0.089$             &$0.033$          &$0.03$         &$0.019$          &$0.033$        \\ \hline
 \multirow{4}{*}{$w_0w_a$CDM}
            &$\sigma(\Omega_{\rm m})$       &$0.035$        &$0.052$             &$0.019$          &$0.029$        &$0.012$          &$0.0078$       \\
            &$\sigma(H_0)$                  &$2.1$          &$1.7$             &$1.2$            &$2.4$          &$0.84$           &$0.83$         \\
            &$\sigma(w_0)$                  &$0.29$         &$0.21$             &$0.13$           &$0.22$         &$0.085$          &$0.082$        \\
            &$\sigma(w_a)$                  &$3.5$          &$1.4$             &$0.57$           &$0.71$         &$0.32$           &$0.32$         \\  \hline
\end{tabular}}}
\end{center}
\end{table*}

In the top-left panel of Fig.~\ref{fig:figure}, we show the constraints on $\Lambda$CDM in the $\Omega_{\rm m}$--$H_0$ plane. As can be seen, HIRAX contributes the most to the FSH results, followed by SKA1-MID\,(0.35\,--\,0.8) and FAST. The combination of them offers $\sigma(\Omega_{\rm m})=0.0039$ and $\sigma(H_0)=0.27\ \rm km\ s^{-1}\ Mpc^{-1}$, which are $(0.0044-0.0039)/0.0044=11\%$ and $(0.32-0.27)/0.32=16\%$ better than those of HIRAX alone. Significantly, FSH (or even HIRAX alone) can place tighter constraints on $\Omega_{\rm m}$ and $H_0$ than CBS. We also compare the constraint results of SKA1-MID and HIRAX in the fifth and sixth columns of Table~\ref{tab:result}. It can be seen that although SKA1-MID has a larger redshift coverage, HIRAX performs better than it in constraining the $\Lambda$CDM model due to the high-precision measurements at $0.8<z<2.5$ (see Fig.~\ref{DAHzfs8}). In general, our joint survey strategy can constrain the $\Lambda$CMD model well, mainly due to the excellent performance of HIRAX.

In the $\Omega_{\rm m}$--$w$ and $H_0$--$w$ planes, we show the constraint results for the $w$CDM model. As can be seen, the contours of SKA1-MID\,(0.35\,--\,0.8) and HIRAX show obviously different parameter degeneracy orientations, especially in the $\Omega_{\rm m}$--$w$ plane. The prime cause is that SKA1-MID\,(0.35\,--\,0.8) and HIRAX are used to observe different epochs of the universe, so they have different sensitivities to the dark-energy EoS. Because HIRAX still dominates the constraints, the synergy of three experiments has limited ability to break the parameter degeneracies. Even so, FSH gives $\sigma(\Omega_{\rm m})=0.0043$, $\sigma(H_0)=0.43\ \rm km\ s^{-1}\ Mpc^{-1}$, and $\sigma(w)=0.019$, which are $12\%$, $26\%$, and $37\%$ better than those of HIRAX alone. In addition, FSH (or even HIRAX alone) could offer tighter constraints on $\Omega_{\rm m}$, $H_0$, and $w$ than CBS, as shown in Table~\ref{tab:result}. We note that SKA1-MID puts a similar constraint on the parameter $w$ as HIRAX, which shows its advantage in constraining the dynamical dark energy model, and the advantage is apparently guaranteed by its survey in the redshift interval $0.35<z<0.8$.

In the bottom-right panel of Fig.~\ref{fig:figure}, we show the constraints on $w_0$ and $w_a$ of the $w_0w_a$CDM model. As can be seen, the contours of SKA1-MID\,(0.35\,--\,0.8) and HIRAX also show different degeneracy orientations. Importantly, their combination can effectively break the degeneracies and thus significantly improve the cosmological parameter constraints. For instance, FSH offers $\sigma(w_0)=0.085$ and $\sigma(w_a)=0.32$, which are $61\%$ and $55\%$ better than those of HIRAX alone. In addition, the joint constraints are almost the same as the results of $\sigma(w_0)=0.082$ and $\sigma(w_a)=0.32$ provided by CBS. We note that depending on the survey at $0.35<z<0.8$, SKA1-MID can put tighter constrains on $w_0$ and $w_a$ than HIRAX, as shown in the fifth and sixth columns of Table~\ref{tab:result}. Therefore, combining FAST, SKA1-MID\,(0.35\,--\,0.8), and HIRAX to constrain cosmological parameters is an excellent choice.

In this paper, we focus on measuring the expansion history of the universe and dark-energy parameters. Note that the 21 cm IM surveys can also be used to measure the structure growth of the universe. We can derive the growth rate $f(z)\simeq\Omega_{\rm m}^{\gamma}(z)$ from RSDs. One may wonder what constraint the joint observation could place on the growth index $\gamma$, which is an important parameter related to linear perturbation. Now we re-perform the MCMC analysis and include $\gamma$ in the parameter sampling. We find that FSH could provide $\sigma(\gamma)=0.013$, $\sigma(\Omega_{\rm m})=0.0042$, and $\sigma(H_0)=0.29\ \rm km\ s^{-1}\ Mpc^{-1}$ in the $\Lambda$CDM model. Furthermore, FSH gives $\sigma(\gamma)=0.022$ in the $w$CDM model and $\sigma(\gamma)=0.023$ in the $w_0w_a$CDM model. Therefore, in any models we consider, the relative error of $\gamma$ is as low as several percent.

\begin{figure*}[!htbp]
\includegraphics[scale=0.37]{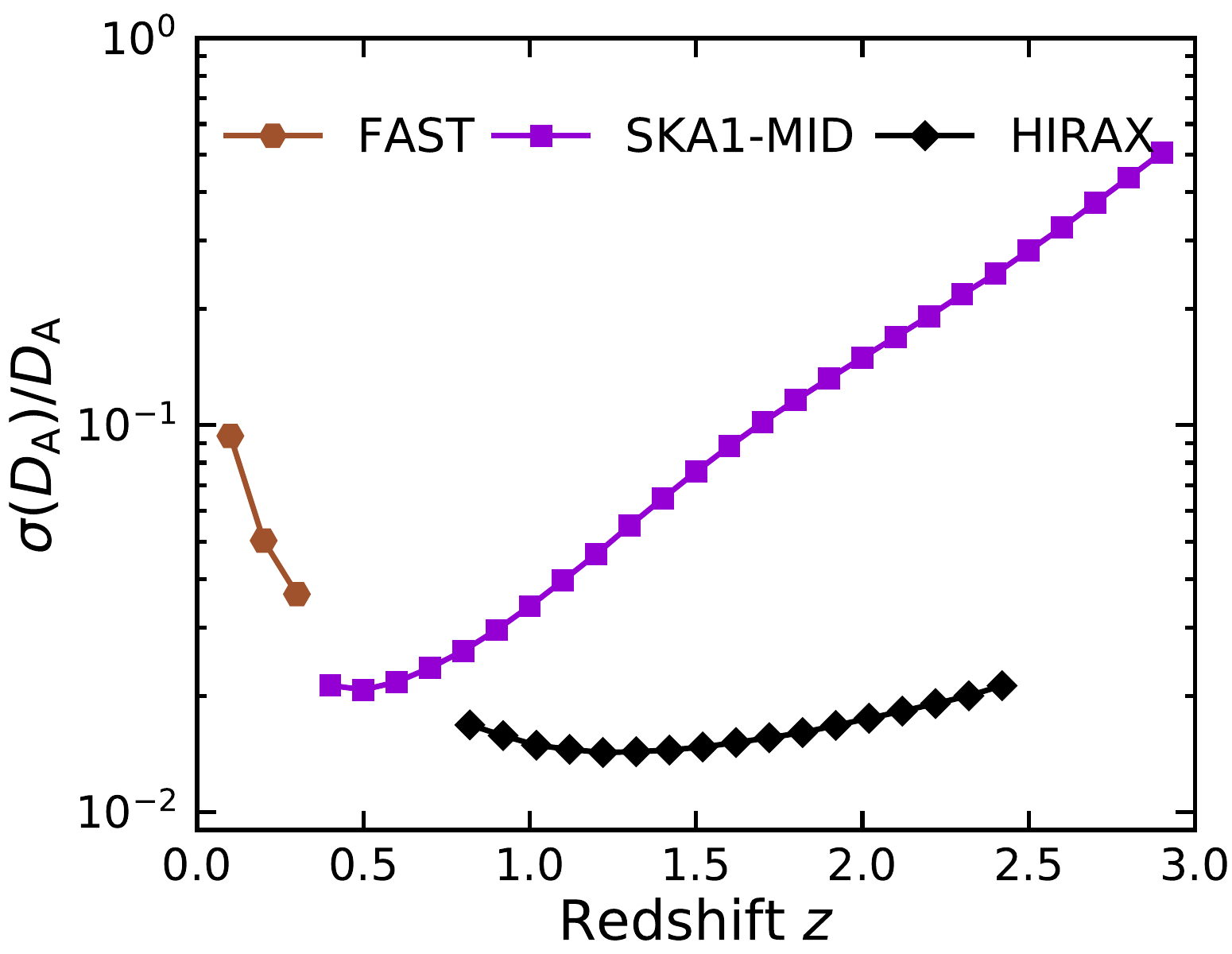}
\includegraphics[scale=0.37]{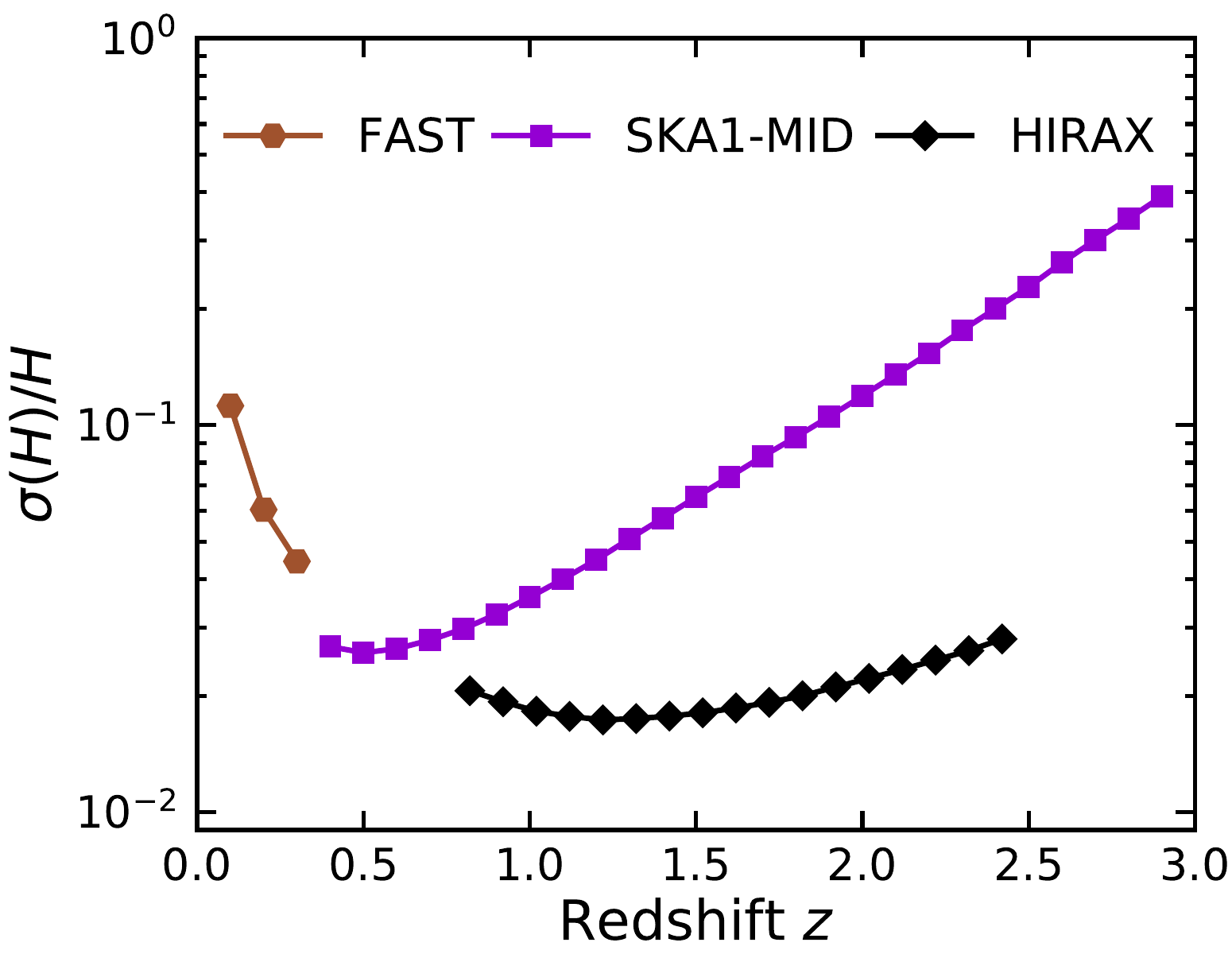}
\includegraphics[scale=0.37]{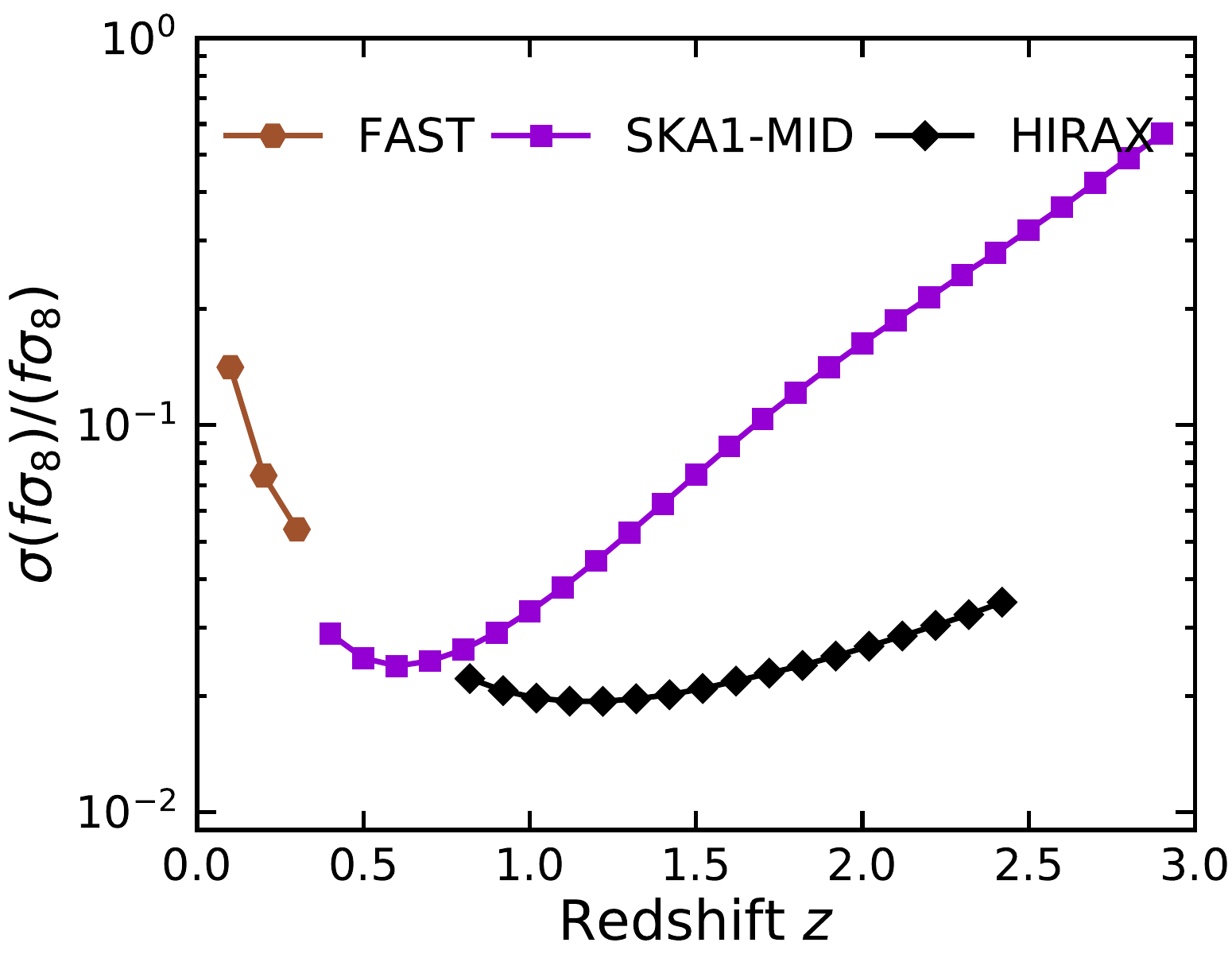}
\caption{Measurement errors on $D_{\rm A}(z)$ (left panel), $H(z)$ (central panel), and $[f\sigma_8](z)$ (right panel) of FAST, SKA1-MID, and HIRAX, in the case of $\varepsilon_{\rm FG}=10^{-5}$.}
\centering
\label{DAHzfs81e5}
\end{figure*}

\begin{figure}
\includegraphics[scale=0.38]{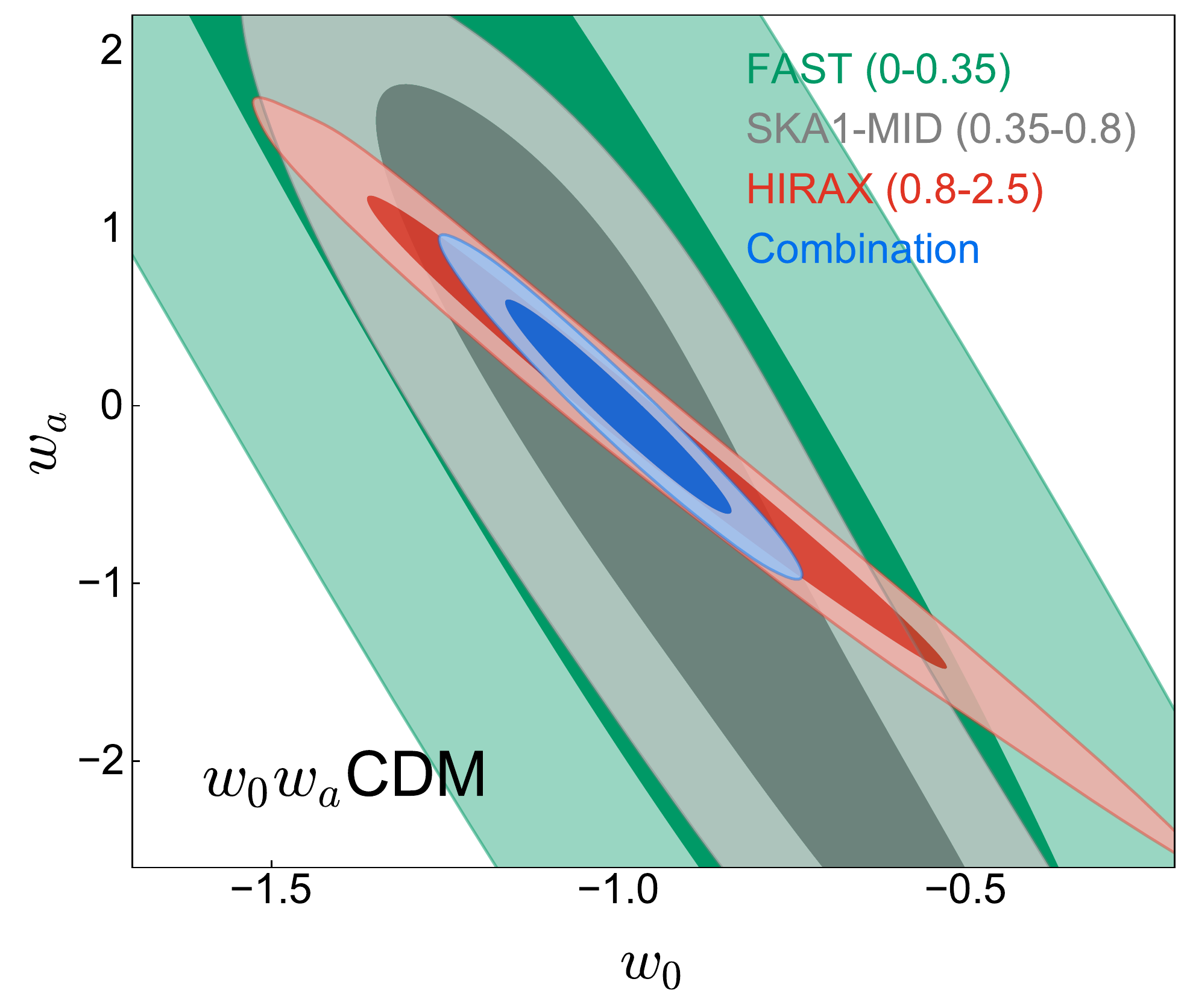}
\centering
\caption{Constraints (68.3\% and 95.4\% confidence level) on $w_0$ and $w_a$ of the $w_0w_a$CDM model by using the FAST, SKA1-MID\,(0.35\,--\,0.8), HIRAX, and FAST+SKA1-MID\,(0.35\,--\,0.8)+HIRAX data, in the case of $\varepsilon_{\rm FG}=10^{-5}$.}
\label{fig:figure1}
\end{figure}

In the above analysis, we have assumed an optimistic scenario for the foreground subtraction, i.e., $\varepsilon_{\rm FG}=10^{-6}$. Now we turn to quantify the sensitivity of our results to the residual foreground contamination amplitude. We re-simulate the 21 cm IM data with a lower foreground removal efficiency of $\varepsilon_{\rm FG}=10^{-5}$, and the measurement errors on $D_{\rm A}(z)$, $H(z)$, and $[f\sigma_8](z)$ are shown in Fig.~\ref{DAHzfs81e5}. As can be seen, the errors of SKA1-MID grow rapidly with redshift and even exceed $40\%$ at $z=3$, but fortunately, they are relatively small at $0.35<z<0.8$. In addition, HIRAX is less affected by the enhanced residual foreground. We use the newly simulated data to constrain the $w_0w_a$CDM model (in which our strategy shows great superiority), and the results are shown in Fig.~\ref{fig:figure1}. It can be seen that the constraints placed by each experiment have become worse. However, the contours of SKA1-MID\,(0.35\,--\,0.8) and HIRAX still show different degeneracy orientations, and thus their combination can effectively break the degeneracies. Concretely, FSH here can provide $\sigma(w_0)=0.11$ and $\sigma(w_a)=0.40$. Although FSH here is inferior to CBS, its constraints on $w_0$ and $w_a$ are $(0.26-0.11)/0.26=58\%$ and $(0.75-0.40)/0.75=47\%$ better than the results of $\sigma(w_0)=0.26$ and $\sigma(w_a)=0.75$ provided by CMB+BAO respectively. To sum up, our strategy performs well in the relatively strong residual foreground.

All previous analyses have assumed specific parameterizations for the dark-energy EoS. To verify the robustness of the conclusion that the joint observation could break the parameter degeneracies, we need to constrain a model with few assumptions. The cosmographic approach meets our purpose \citep{Capozziello:2017ddd,Li:2019qic,Capozziello:2019cav,Mandal:2020buf, Rezaei:2020lfy, Rezaei:2021qwd, Pourojaghi:2022zrh, Mu:2023bsf}. In this approach, the higher-order Taylor expansion for $H(z)$ can be written as \citep{Capozziello:2017ddd}
\begin{align}
H(z) &\simeq H_{0}\bigg[1+z\left(1+q_{0}\right)+\frac{z^{2}}{2}\left(j_{0}-q_{0}^{2}\right) \nonumber \\
&+\frac{z^{3}}{6}\left(-3 q_{0}^{2}-3 q_{0}^{3}+j_{0}\left(3+4 q_{0}\right)+s_{0}\right)\bigg],
\end{align}
where the cosmographic parameters (Hubble, deceleration, jerk and snap) are defined as \citep{Visser:2003vq}
\begin{align}
\begin{aligned}
H & \equiv \frac{1}{a} \frac{d a}{d t},  &q & \equiv-\frac{1}{a H^{2}} \frac{d^{2} a}{d t^{2}}, \\
j & \equiv \frac{1}{a H^{3}} \frac{d^{3} a}{d t^{3}}, &s & \equiv \frac{1}{a H^{4}} \frac{d^{4} a}{d t^{4}},
\end{aligned}
\end{align}
and $H_0$, $q_0$, $j_0$ and $s_0$ represent their values today. Note that the expansion is not accurate at $z>1$. Therefore, it is not appropriate to use the FSH data (which covers $0<z<2.5$) to constrain the model. But just to study the performance of our strategy, we still use the simulated data to constrain the cosmographic parameters. The results are shown in Table~\ref{tab:result1} and Fig.~\ref{fig:figure1}. As can be seen, the contours of SKA1-MID\,(0.35\,--\,0.8) and HIRAX show different degeneracy orientations, so their combination can break the degeneracies and thus improve the constraints. Specifically, FSH provides $\sigma(H_0)=0.22\ \rm km\ s^{-1}\ Mpc^{-1}$, $\sigma(q_0)=0.024$, $\sigma(j_0)=0.077$ and $\sigma(s_0)=0.083$, which are $48\%$, $40\%$, $36\%$ and $36\%$ better than those given by HIRAX, respectively. In addition, FSH could place tighter constraint on $H_0$ here than in $\Lambda$CDM. The results prove the value of our strategy again. We stress that the Taylor expansion is only valid at low redshifts, so the results are only for reference.

\begin{table}[!htb]
\caption{The 1$\sigma$ errors on cosmographic parameters by using the FAST, SKA1-MID\,(0.35\,--\,0.8), SKA1-MID, HIRAX, and FSH data. Here $H_0$ is in units of $\rm km\ s^{-1}\ Mpc^{-1}$.}
\label{tab:result1}
\setlength{\tabcolsep}{0.2mm}
\renewcommand{\arraystretch}{1.2}
\begin{center}{\centerline{
\begin{tabular}{ccm{1.3cm}<{\centering}m{1.7cm}<{\centering}m{1.7cm}<{\centering}m{1.4cm}<{\centering}m{1.3cm}<{\centering}}
\hline
            & Error                 &FAST           &SKA1-MID (0.35--0.8)   &SKA1-MID           &HIRAX              &FSH                    \\ \hline
 \multirow{4}{*}
            &$\sigma(H_0)$          &$0.90$         &$0.56$                 &$0.30$             &$0.42$             &$0.22$               \\
            &$\sigma(q_0)$          &$0.18$         &$0.14$                 &$0.034$            &$0.040$            &$0.024$                 \\
            &$\sigma(j_0)$          &$-$            &$0.95$                 &$0.12$             &$0.12$             &$0.077$               \\
            &$\sigma(s_0)$          &$-$            &$1.5$                  &$0.12$             &$0.13$             &$0.083$                \\ \hline
\end{tabular}}}
\end{center}
\end{table}

\begin{figure*}
\includegraphics[scale=0.5]{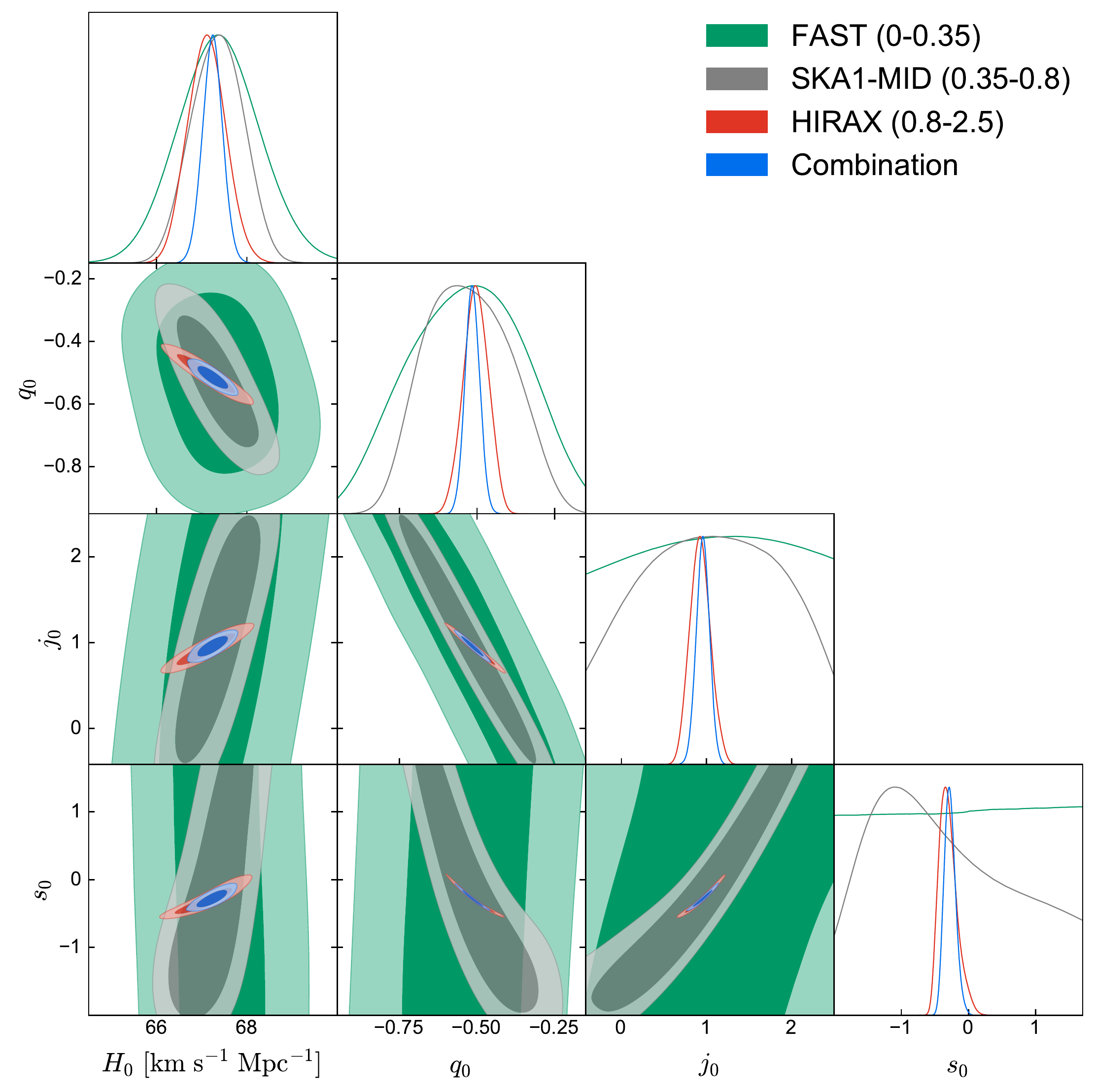}
\centering
\caption{Constraints (68.3\% and 95.4\% confidence level) on cosmographic parameters by using the FAST, SKA1-MID\,(0.35\,--\,0.8), HIRAX, and FAST+SKA1-MID\,(0.35\,--\,0.8)+HIRAX data, in the case of $\varepsilon_{\rm FG}=10^{-6}$.}
\label{fig:figure1}
\end{figure*}

Our results are sufficient to show that FAST+SKA1-MID\,(0.35\,--\,0.8)+HIRAX is a promising 21 cm IM survey strategy. In the strategy, FAST observes the dark energy-dominated epoch of the universe, SKA1-MID\,(0.35\,--\,0.8) covers the transition redshift that determines the onset of cosmic acceleration, and HIRAX mainly measures the matter-dominated era of the universe. Due to the precise measurements of BAOs and RSDs at $0.8<z<2.5$, HIRAX has a great potential in constraining the dark energy model with a simple EoS, such as the $\Lambda$CDM and $w$CDM models. However, it performs not well in constraining the dynamical dark energy model having a more sophisticatedly parameterized EoS, such as the $w_0w_a$CDM model. Fortunately, SKA1-MID\,(0.35\,--\,0.8) has different parameter degeneracy orientations from HIRAX, so their combination can effectively break the degeneracies. It should be pointed out that although FAST contributes the least to SFH results, it is also useful for improving the cosmological parameter estimation. In conclusion, the 21 cm IM joint survey strategy proposed here is worth pursuing.

\section{Conclusions}\label{sec4}
In this work, we explore the potential of using a novel 21 cm IM joint survey strategy, FAST\,($0<z<0.35$)\,+\,SKA1-MID\,($0.35<z<0.8$)\,+\,HIRAX\,($0.8<z<2.5$), to measure dark energy. The strategy could give full play to the advantages of relevant experiments. We simulated the 21 cm IM observations under the assumption of excellent foreground removal, and used the mock data to constrain three typical dark energy models, i.e., the $\Lambda$CDM, $w$CDM, and $w_0w_a$CDM models.

We find that the synergy of three experiments could provide tight constraints on cosmological parameters. For instance, it offers $\sigma(\Omega_{\rm m})=0.0039$ and $\sigma(H_0)=0.27\ \rm km\ s^{-1}\ Mpc^{-1}$ in the $\Lambda$CDM model, $\sigma(w)=0.019$ in the $w$CDM model, and $\sigma(w_0)=0.085$ and $\sigma(w_a)=0.32$ in the $w_0w_a$CDM model. The constraint results are better than or at least equal to those given by the combination of three mainstream observations, CMB+BAO+SN. In addition, we find that as the parameterization of the dark energy EoS becomes more complex, our strategy gradually demonstrates its superiority. For instance, the joint observation gives $\sigma(w_0)=0.085$ and $\sigma(w_a)=0.32$, which are $61\%$ and $55\%$ better than those of HIRAX alone, due to the parameter degeneracies being broken. All of these show that the 21 cm IM joint survey strategy is promising and worth pursuing.

We prove the robustness of our conclusion that the joint observation can break the degeneracies by constraining the cosmographic parameters. The joint observation gives $\sigma(H_0)=0.22\ \rm km\ s^{-1}\ Mpc^{-1}$, $\sigma(q_0)=0.024$, $\sigma(j_0)=0.077$ and $\sigma(s_0)=0.083$, which are significantly better than those given by HIRAX alone.

Note that we do not consider the survey of SKA1-MID at $2.5<z<3.05$ in the strategy, due to the relatively poor-quality measurements in that region (as shown in Fig.~\ref{DAHzfs8}). If the survey at $2.5<z<3.05$ is included, we can further improve the cosmological parameter constraints. In addition, it should be pointed out that in the strategy, HIRAX can be replaced by CHIME or Tianlai (full-scale experiment), because they can also precisely measure the BAOs and RSDs in the redshift interval $0.8<z<2.5$ \citep{Wu:2021vfz}.

In this work, we assumed that the foreground of 21\,cm\,IM measurements can be subtracted to an extremely low level. We have tested that the survey strategy still performs well in a relatively strong residual foreground.

\begin{acknowledgments}
This work was supported by
the National SKA Program of China (Grants Nos. 2022SKA0110200 and 2022SKA0110203) and
the National Natural Science Foundation of China (Grants Nos. 11975072, 11835009, and 11875102).
\end{acknowledgments}

\bibliography{21cmIM}

\end{document}